\documentclass[12pt]{iopart}
\begin{document}
\input{epsf.sty}
\jl{4}
\title[Hadronic B decays]{Factorization and hadronic $B$ decays 
in the heavy quark limit}

\author{W N Cottingham\dag, I B Whittingham\ddag,
         N de Groot\S,
        and F Wilson\dag}

\address{\dag\ H H Wills Physics Laboratory, 
Royal Fort, Tyndall Ave, Bristol, BS8~1TL, UK}

\address{\ddag\ School of Mathematical and Physical Sciences, 
James Cook University, Townsville, Australia, 4811}

\address{\S\ Rutherford Appleton Laboratory,
Chilton, Didcot OX11 0QX, UK}

\begin{abstract}
Recent theoretical investigations (M. Beneke \etal 1999 
{\it Phys. Rev. Lett.} {\bf 83} 1914 ) of two-body hadronic 
$B$ decays have provided justification, at least to order $\alpha_{s}$, 
for the use of the factorization ansatz to evaluate the $B$ decay 
matrix elements provided the decay meson containing the spectator quark 
is not heavy. Although data is now available on many charmless two-body $B$
decay channels, it is so far not very precise with systematic and
statistical
errors in total of the order of 25\% or greater. Analyses have been made on
$\pi \pi$ and $\pi K$ channels with somewhat contradictory results (M. Beneke
\etal 2001 {\it Nucl. Phys. B} {\bf 606} 245; M. Ciuchini \etal 2001
hep-ph/0110022). We take an overview of these channels and of other
channels containing vector mesons.
Because factorization involves many poorly known soft QCD parameters, and
because of the imprecision of current data, we present simplified
formulae for a wide range of $B$ decays into the lowest mass pseudoscalar
and vector mesons.  These formulae, valid in the heavy quark limit, involve
a reduced set of soft QCD parameters and, although resulting in some loss
of accuracy, should still provide an adequate and transparent tool 
with which to confront data for some time to come.
Finally we confront these formulae with data on nineteen channels. We find
a plausible set of soft QCD parameters that, apart from three pseudoscalar 
vector channels, fit the branching ratios and the recently measured value of
$\sin (2\beta)$ quite well.\end{abstract}


\section{Introduction}
Much experimental effort is being expended in the study of $B$ meson
decays \cite{CLEO,Babar,BaBarbeta,Ba0111,Belle,Belle2001}
and the next decade will see
intensive investigation of the $B$-meson system at the Tevatron, 
the SLAC and KEK $B$-factories, and at the LHC.
The aim is to establish the Cabbibo-Kobayashi-Maskawa (CKM) parameters to
an accuracy that will test the consistency of the Standard Model (SM)
description of CP asymmetry, hopefully to a precision comparable to that
of other aspects of the Standard Model. $B$ meson decays should have
many channels which exhibit CP asymmetry but even with precise data there  
will be a problem of reliably unravelling the underlying 
weak decay mechanisms from the distortions caused by the strong 
interactions.

Hard QCD corrections to the underlying $b$ quark weak decay amplitudes
involve gluon virtualities between the electroweak scale $M_{W}$ and
the scale $\Or(m_{b})$ and are implemented through
renormalization of the calculable short distance Wilson coefficients
$C_{i}$ in the low energy effective weak Hamiltonian\cite{Buch96} 
for $\Delta B =1$ decays at scale $\mu =\Or(m_{b})$ 
\begin{equation}
\label{fact1}
\fl
H_{{\rm eff}}(\mu ) =  \frac{G_{F}}{\sqrt{2}}
\left\{\sum_{p=u,c} \lambda_{p} \;[C_{1}(\mu )O^{p}_{1}
+C_{2}(\mu )O^{p}_{2}]  
-\lambda_{t} \sum_{i=3, \ldots ,6}\;C_{i}(\mu )O_{i} \right\}   
 + {\rm other\; terms}
\end{equation}
where  $\lambda_{p} \equiv V^{*}_{pq}V_{pb}$ is a product of CKM matrix
elements, $q=d,s$ and the local $\Delta B =1$ four-quark operators are
\begin{eqnarray}
\label{fact2}
O_{1}^{p} & \equiv & (\bar{q}_{\alpha}\;p_{\alpha})_{{\rm V-A}}\;
(\bar{p}_{\beta}\;b_{\beta})_{{\rm V-A}}, \nonumber \\
O_{2}^{p} & \equiv & (\bar{q}_{\alpha}\;p_{\beta})_{{\rm V-A}}\;
(\bar{p}_{\beta}\;b_{\alpha})_{{\rm V-A}}, \nonumber  \\
O_{3,5}& \equiv & (\bar{q}_{\alpha}\;b_{\alpha})_{{\rm V-A}}\;
\sum_{q^{\prime}}(\bar{q}^{\prime}_{\beta}\;
q^{\prime}_{\beta})_{{\rm V} \mp {\rm A}}  \nonumber \\
O_{4,6}& \equiv & (\bar{q}_{\alpha}\;b_{\beta})_{{\rm V-A}}\;
\sum_{q^{\prime}}(\bar{q}^{\prime}_{\beta}\;
q^{\prime}_{\alpha})_{{\rm V} \mp {\rm A}},  
\end{eqnarray}
where $q^{\prime} \in \{u,d,s,c\}$, 
$\alpha$ and $\beta$ are colour indices, and we have used the notation, 
for example,
\begin{equation}
\label{fact3}
(\bar{q}_{\alpha}\;b_{\alpha})_{{\rm V-A}}\;(\bar{q}^{\prime}_{\beta}\;
q^{\prime}_{\beta})_{{\rm V} \mp {\rm A}}
  = [\bar{q}_{\alpha}\gamma_{\mu}(1-\gamma_{5})b_{\alpha}] 
 [\bar{q}^{\prime}_{\beta}\gamma^{\mu}(1 \mp \gamma_{5})q^{\prime}_{\beta}].
\end{equation}
The operators in (\ref{fact2}) are associated with particular processes
(see \cite{Buch96} for a detailed discussion):
$O_{1,2}$ are the tree current-current operators and
$O_{3, \ldots ,6}$ are QCD penguin operators. The ``other terms" 
indicated in (\ref{fact1}) are small in the SM. They include magnetic 
dipole transition operators and electroweak penguin operators.
The most important of these is the electroweak penguin operator
\begin{equation}
\label{fact3a}
O_{9} \equiv \frac{3}{2} (\bar{q}_{\alpha}\;b_{\alpha})_{{\rm V-A}}\;
\sum_{q^{\prime}}e_{q^{\prime}}(\bar{q}^{\prime}_{\beta}\;
q^{\prime}_{\beta})_{{\rm V} - {\rm A}}
\end{equation}
where $e_{q^{\prime}} = 2/3\;(-1/3)$ for $u(d)$ -type quarks. The Wilson
coefficient $C_{9}$ is larger than the QCD penguin coefficients $C_{3}$
and $C_{5}$ and contributes $-(G_{F}/\sqrt{2})\lambda_{t}C_{9}(\mu )O_{9}$
to $H_{{\rm eff}}(\mu )$.

Inclusion of strong interaction effects below the scale $\mu $ is a very
difficult task involving, for the two-body hadronic decay $B \rightarrow
h_{1}h_{2}$, the computation of the matrix elements
$\langle h_{1}h_{2}|O_{i}|B\rangle $. 
Until recently, most theoretical 
studies of two-body hadronic decay invoked the factorization 
approximation \cite{factor} in which final state interactions are
neglected and $\langle h_{1}h_{2}|O_{i}|B\rangle $ is expressed as a
product of two hadronic currents: 
$\langle h_{1}|J_{1\;\mu}|B \rangle \langle h_{2}| J^{\mu }_{2}|0\rangle $.
The operators $O^{p}_{2}$ and $O_{4,6}$ are Fierz transformed into a 
combination of colour singlet-singlet
and octet-octet terms and the octet-octet terms then discarded. 
The singlet-singlet current matrix elements are then
expressed in terms of known decay rates and form factors. 
Consequently, the hadronic matrix elements are expressed in terms of 
the combinations
\begin{equation}
\label{fact4}
a_{2i-1}=C_{2i-1}+\frac{1}{N_{c}}C_{2i}, \qquad
a_{2i}=C_{2i}+\frac{1}{N_{c}}C_{2i-1}
\end{equation}
where $i=1,2,3$ and $N_{c}=3$ is the number of colours.

In the widely used so-called ``generalized factorization" approach 
\cite{AliGreub98,Alietal98,Cheng98,ChengYang99}, 
the renormalization scale dependence of the hadronic 
matrix elements $\langle h_{1}h_{2}|O_{i}|B\rangle $, lost through 
factorization, is compensated for through the introduction of 
effective Wilson coefficients $C^{{\rm eff}}_{i}(\mu )$ such that
\begin{equation}
\label{fact5}
C_{i}(\mu )\langle O_{i}(\mu ) \rangle = C_{i}^{{\rm eff}}(\mu )
\langle O_{i} \rangle ^{{\rm tree}}.
\end{equation}
The effective Wilson coefficients $C^{{\rm eff}}_{i}(\mu ), i=3, \ldots, 6$
for the QCD penguins depend upon the gluon momentum $q^{2}$ and generate
strong phases as $q^{2}$ crosses the $u\bar{u}$ and $c\bar{c}$ 
thresholds \cite{Band79}. The neglected octet-octet terms are compensated
for by replacing $N_{c}$ by a universal free parameter $\xi $.
The assumed universal $a_{i}$ parameters are then determined by fitting
to as much data as possible. Some authors \cite{ChengYang99} 
have allowed the $\xi $
parameter for the $(V-A) \otimes (V-A)$ and $(V-A) \otimes (V+A)$ 
contributions to be different.

Recently there has been significant progress in the theoretical
understanding of hadronic decay amplitudes in the heavy quark 
limit 
\cite{Beneke99,Benekeetal2000,Beneke2000,Neubert2000,Du2000,Yang2000,
Muta2000,ChengYang2001a,ChengYang2001b,Benekeetal2001,Duetal2002}.
These approaches, known as 
QCD (improved) factorization, exploit the fact that $m_{b}$ is much 
greater than the QCD scale $\Lambda_{{\rm QCD}}$ and show that the
hadronic matrix elements have the form
\begin{equation}
\label{fact6}
\langle h_{1}h_{2}|O_{i}|B \rangle  =  
\langle h_{1}|J_{1\;\mu}|B \rangle 
\langle h_{2}|J_{2}^{\mu}|0 \rangle  
 \left[1+\sum_{n} r_{n}\alpha_{s}^{n} + \Or(\Lambda_{{\rm QCD}}/m_{b})\right]
\end{equation}
provided the spectator quark does not go to a heavy meson. If the power
corrections in $\Lambda_{{\rm QCD}}$ and radiative corrections in 
$\alpha_{s}$ are neglected, conventional or ``naive" factorization is
recovered. Although naive factorization is broken at higher order in 
$\alpha_{s}$, these non-factorizable contributions can be calculated
systematically.

For $B \rightarrow h_{1}h_{2}$ in which both $h_{1}$ and $h_{2}$ are light,
and $h_{2}$ is the meson that does not pick up the spectator quark, 
\cite{Beneke99,Benekeetal2000,Benekeetal2001} 
find that all non-factorizable contributions are real
and dominated by hard gluon exchange which can be calculated perturbatively,
and all leading order non-perturbative soft and collinear effects are 
confined to the $B-h_{1}$ system and can be absorbed into form factors
and light cone distribution amplitudes. Strong rescattering  phases are
either perturbative or power suppressed in $m_{b}$ and, at leading order,
arise
via the Bander-Silverman-Soni mechanism \cite{BBS} 
from the imaginary parts of the hard scattering kernels in $r_{1}$.
The contribution from the annihilation diagram (in which the spectator
quark annihilates with one of the $b$ decay quarks) is found to be
power suppressed. In contrast to one of the basic assumptions
of generalized factorization, the corrections to naive factorization are
process dependent and have a richer structure than merely allowing $\xi $
to be different for the  $(V-A)\otimes (V-A)$ and $(V-A)\otimes (V+A)$
contributions.

Similar in many ways to QCD factorization is the hard scattering or
perturbative QCD
approach \cite{hard}. However, here it is argued that all soft contributions
to the $B-h_{1}$ form factor are negligible
if transverse momentum with Sudakov resummation is included,
so that the form factors can
be perturbatively calculated. Also, it is argued that Sudakov
suppression of long distance effects in the $B$ meson is needed to 
control higher order effects in the 
Beneke {\it et al} \cite{Beneke99,Benekeetal2000} approach.
Non-factorizable contributions are now found to be complex
but the imaginary parts are much smaller than that of the (factorizable)
contribution from the annihilation diagram.
Several aspects of the perturbative QCD approach have been strongly
criticized \cite{Benekeetal2001,Quinn2001}.

In summary, recent theoretical investigations have provided justification,
at least to order $\alpha_{s}$, for invoking factorization to evaluate
the hadronic matrix elements $\langle h_{1}h_{2}|O_{i}|B \rangle $ 
provided the decay meson containing the spectator quark is not heavy.
Unfortunately, the numerical results from the QCD factorization
calculations are quite sensitive to the input assumptions on the quark
distribution functions \cite{Quinn2001b},
and the contributions from the logarithmic end point divergences in hard
spectator scatterings and weak annihilation \cite{Duetal2002},
thus giving rise to significant
theoretical uncertainties which currently limit their usefulness.

The success claimed \cite{Benekeetal2001} for QCD improved factorization in
fitting the observed branching ratios for $B \rightarrow \pi \pi $ and
$B \rightarrow K \pi $ has been queried by Ciuchini {\it et al}
\cite{Ciuchini2001} who argue that, with the CKM angle $\gamma $
constrained by the unitarity triangle analysis of \cite{Ciuchini2000},
non-perturbative
$\Lambda_{{\rm QCD}}/m_{b}$ corrections, so called ``charming penguins",
are important where the factorized amplitudes are either colour or
Cabbibo suppressed. Factorization at leading order in
$\Lambda_{{\rm QCD}}/m_{b}$ cannot reproduce the observed
$B \rightarrow \pi \pi$ and $B \rightarrow K \pi $ decays and must be
supplemented with enhanced $c$-loop penguin diagrams. 

In this paper we present formulae based on the factorization ansatz  
for $B$ decay into the lowest mass pseudoscalar and vector mesons. 
We focus on the heavy meson limit in 
which combinations of the different parameters that characterize the 
current matrix elements reduce to just one. This is a consistent approach
within the heavy quark approximation and we believe that the resulting
more simple formulae will provide an adequate and more transparent tool
with which to confront data for some time to come.
Finally we confront these formulae with data on nineteen channels.
With $\sin (2\beta)$ as measured by the
BaBar \cite{BaBarbeta} and Belle collaborations \cite{Belle2001} we find
a plausible set of soft QCD parameters that, apart from three pseudoscalar 
vector channels, fit the data well.

\section{Factorized decay amplitude}

The amplitude for charmless hadronic $b$ decays 
for all these factorization schemes can be written as
\begin{equation}
\label{fact7}
\langle h_{1}h_{2}|H^{{\rm eff}}|B \rangle = \frac{G_{F}}{\sqrt{2}}
\sum_{i=1,\ldots, 6,9} p_{i}[Q_{i}(h_{1},h_{2}) + Q_{i}(h_{2},h_{1})]
\end{equation}
where
\begin{eqnarray}
\label{fact8}
Q_{1}(h_{1},h_{2})&=& \langle h_{1}|(\bar{u}_{\alpha}\;
b_{\alpha})_{{\rm V-A}}|B \rangle \; \langle h_{2}|(\bar{q}_{\beta}\;
u_{\beta})_{{\rm V-A}}|0 \rangle , \nonumber  \\
Q_{2}(h_{1},h_{2})&=& \langle h_{1}|(\bar{q}_{\alpha}\;
b_{\alpha})_{{\rm V-A}}|B \rangle \; \langle h_{2}|(\bar{u}_{\beta}\;
u_{\beta})_{{\rm V-A}}|0 \rangle , \nonumber  \\
Q_{3}(h_{1},h_{2})&=& \langle h_{1}|(\bar{q}_{\alpha}\;
b_{\alpha})_{{\rm V-A}}|B \rangle \; 
\langle h_{2}|(\bar{q}^{\prime}_{\beta}\;
q^{\prime}_{\beta})_{{\rm V-A}}|0 \rangle , \nonumber  \\
Q_{4}(h_{1},h_{2})&=& \langle h_{1}|(\bar{q}^{\prime}_{\alpha}\;
b_{\alpha})_{{\rm V-A}}|B \rangle \; 
\langle h_{2}|(\bar{q}_{\beta}\; 
q^{\prime}_{\beta})_{{\rm V-A}}|0 \rangle , \nonumber  \\
Q_{5}(h_{1},h_{2})&=& \langle h_{1}|(\bar{q}_{\alpha}\;
b_{\alpha})_{{\rm V-A}}|B \rangle \; 
\langle h_{2}|(\bar{q}^{\prime}_{\beta}\;
q^{\prime}_{\beta})_{{\rm V+A}}|0 \rangle , \nonumber  \\
Q_{6}(h_{1},h_{2})&=&-2 \langle h_{1}|(\bar{q}^{\prime}_{\alpha}\;
b_{\alpha})_{{\rm S-P}}|B \rangle \; 
\langle h_{2}|(\bar{q}_{\beta}\;
q^{\prime}_{\beta})_{{\rm S+P}}|0 \rangle , \nonumber  \\
Q_{9}(h_{1},h_{2})&=& \langle h_{1}|(\bar{q}_{\alpha}\;
b_{\alpha})_{{\rm V-A}}|B \rangle \; 
\langle h_{2}|e_{q^{\prime}}(\bar{q}^{\prime}_{\beta}\;
q^{\prime}_{\beta})_{{\rm V-A}}|0 \rangle , 
\end{eqnarray}
with
\begin{equation}
\label{fact9}
(\bar{q}^{\prime}_{\alpha}\;b_{\alpha})_{{\rm S-P}}  =  
\bar{q}^{\prime}_{\alpha} (1-\gamma_{5}) b_{\alpha}, \qquad
(\bar{q}_{\beta} q^{\prime}_{\beta})_{{\rm S+P}}  =  
\bar{q}_{\beta} (1+\gamma_{5}) q^{\prime}_{\beta},
\end{equation}
and
\begin{equation}
\label{fact9a}
p_{1,2} \equiv \lambda_{u} a_{1,2}; \qquad 
p_{3,\ldots ,6,9} \equiv  \lambda_{u} a^u_{3, \ldots ,6,9}
\ + \lambda_{c} a^c_{3, \ldots ,6,9}.
\end{equation}
The form of the matrix element $Q_{6}$ is a consequence of a Fierz 
transformation on the $(V-A) \otimes (V+A)$ term.
An example of a process in which $h_{1}$ (and $h_{2}$) can attach 
itself to either current is
for  $\bar{q}^{\prime}=\bar{q}=d$ where the 
matrix element for $\bar{B}^{0} \rightarrow \pi^{0}\rho^{0}$ will have
contributions from both terms.

The matrix elements in $Q_{6}$ can be estimated from the matrix elements
of the electroweak currents by taking the quarks, at some appropriate
mass scale, to be on mass shell. For the current
\begin{equation}
\label{fact10}
J^{\mu} = \bar{q}_{1} \gamma^{\mu}(1 \mp \gamma_{5}) q_{2}
\end{equation}
this yields
\begin{equation}
\label{fact11}
\rmi \partial _{\mu} J^{\mu} = (m_{2}-m_{1})\bar{q}_{1}q_{2} 
\pm (m_{1}+m_{2}) \bar{q}_{1} \gamma_{5} q_{2}.
\end{equation}
Forming matrix elements of (\ref{fact11}) between scalar, pseudoscalar and
vector states as appropriate, one of the terms on the right hand side will 
be identically zero and the other will then be determined by the matrix 
element of the left hand side.

The $a_{i}$ coefficients in (\ref{fact7}) have the form
\begin{equation}
\label{fact12}
a_{i}=a_{i}^{{\rm LO}} + \alpha_{s} a_{i}^{(1)}
\end{equation}
where the leading order (LO) $a_{i}^{{\rm LO}}$ are given by the naive
factorization expressions (\ref{fact4}) with the Wilson coefficients
$C_{i}$ taken at next to leading order (NLO). Detailed expressions for the
$a_{i}^{(1)}$ are given in
\cite{Beneke99,Benekeetal2000,Du2000,Muta2000,ChengYang2001b,Benekeetal2001}
for $B \rightarrow PP$,
in \cite{Yang2000} for $B \rightarrow PV$ and in \cite{ChengYang2001a} for
$B \rightarrow VV$, where $P$ and $V$ denote light
pseudoscalar and vector mesons respectively.
Note that \cite{Benekeetal2001} claim there are errors in the expressions
given by \cite{Du2000,Muta2000,ChengYang2001b} for $B \rightarrow PP$.

The beauty of the result (\ref{fact12}) is that the difference between the
decay rate formulae of naive and generalized factorization as presented
in numerous works \cite{AliGreub98,Alietal98,ChengYang99} and that based
upon QCD factorization lies only in the coefficients $a_{i}$, the 
factorization matrix elements $Q_{i}$ are common to all these approaches.
Also, if the soft gluon physics is all accounted for in the current matrix
elements then, in the heavy quark limit, $N_{c}$ should be taken to be
three.

The corrections to the coefficients $a_{i}$ so far presented are to first
order in $\alpha_{s}$. An encouraging feature
is that they are not generally large,
which leads one to hope that the precision with which the SM can be 
tested will be determined by the proximity of the $b$ quark mass to the 
heavy quark limit and the precision of our knowledge of the soft QCD 
parameters, $B$ meson semileptonic transition form factors and meson light
cone distribution amplitudes. An example of the differences in the $a_{i}$ 
coefficients for several theoretical approaches is shown in Table 1.
Here we compare the $a_{i}$ coefficients  for
QCD factorization, generalized factorization and a simple  
tree plus penguin model at scale $M_{W}$. 
In this simple model, for example, the process
$b \rightarrow d q\bar{q}$ produces a $q \bar{q}$ pair in a colour octet
state so that they do not materialize in a $q \bar{q}$ meson; the $\bar{q}$
must pair with the $d$ quark and the $q$ quark with the spectator antiquark
so that $a_{3}=a_{5}=0$ and $a_{4}=a_{6}$.
We have estimated these coefficients by assuming a quark parton distribution
function  $6x(1-x)$ in both $h_{1}$ and $h_{2}$ and that, for a given
$x_{1}$ and $x_{2}$, the $q\bar{q}$ pair has, neglecting quark masses,
an invariant mass squared $q^{2}=x_{1}x_{2}M_{B}^{2}$. The coefficient
$a_{4}(=a_{6})$ was then calculated by integrating the penguin amplitude
given in \cite{ACW98}.
All estimates of the penguin amplitudes and the associated $a_{i}$
coefficients are uncertain; they involve the strong coupling $\alpha_{s}$
over a distribution in $q^{2}$, models of quarks in hadrons and soft
QCD parameters, some of which are
only poorly known.
Because of these uncertainties, to be confident of any conclusions drawn
about the SM it will be important to have a consistent picture of as many
channels of charmless $B$ decay as possible.

\section{Reduction of factorization matrix elements}

The factorization matrix elements $Q_{i}$ in (\ref{fact8}) involve 
products of current matrix elements which are evaluated through the 
introduction of numerous soft QCD parameters such as meson decay constants
and transition form factors. Explicit expressions given in the literature
\cite{AliGreub98,Alietal98,Yang2000} for $B \rightarrow PP$, $B \rightarrow
PV$ and $B \rightarrow VV$ decay amplitudes in the factorization 
approximation are extremely cumbersome and, consequently, are unlikely in
the short term to be of great assistance to experimentalists in 
analyzing their limited data sets. We argue that the expressions for these 
factorized decay amplitudes can be simplified, albeit at some loss of
accuracy, such that a wide range of decays can be expressed in terms of a
relatively small number of soft QCD parameters which are, or will be, 
relatively well known.
Our approximations consist of (1) neglect of terms $(m_{P,V}/m_{B})^{2}$
in the decay amplitudes and rates and (2) use of universal
generalized-factorization model $a_{i}$ coefficients rather than the
(slightly) process-dependent QCD-factorization model $a_{i}$ coefficients.
The effect of this latter approximation is considered in section 5 for decays
involving $\pi $ and $K$ mesons for which the QCD factorization $a_{i}$
coefficients are well established.
The loss of accuracy incurred in our approach
should not be significant until more precise data is available. Our
expressions are formally exact in the heavy quark limit.

The $\pi $ and $K$ meson decay constants are defined through the matrix
elements
\begin{equation}
\label{fact13}
\langle \pi^{-} | \bar{d}_{\alpha} \gamma_{5} \gamma^{\mu}u_{\alpha}
| 0 \rangle   =  - f_{\pi} p_{\pi}^{\mu}, \qquad 
\langle K^{-} | \bar{s}_{\alpha} \gamma_{5} \gamma^{\mu} u_{\alpha}
| 0 \rangle  =  - f_{K} p_{K}^{\mu}.
\end{equation}
Isospin symmetry then determines that
\begin{equation}
\label{fact14}
\langle \pi^{0} | \bar{u}_{\alpha}\gamma_{5} \gamma^{\mu} u_{\alpha}
| 0 \rangle = - \langle \pi^{0} | \bar{d}_{\alpha} \gamma_{5} \gamma^{\mu}
d_{\alpha} | 0 \rangle = - \frac{1}{\sqrt{2}} f_{\pi} p_{\pi}^{\mu},
\end{equation}
with similar relations for the $K$ meson. 
The magnitudes of $f_{\pi}$ and $f_{K}$ are well
determined from experimental measurements of the leptonic decays such as
\begin{equation}
\label{fact15}
\Gamma(\pi^{-} \rightarrow l^{-}\bar{\nu})= \frac{G_{F}^{2}}{8 \pi}
|V_{ud}|^{2}f_{\pi}^{2}\; m_{\pi}m_{l}^{2}
\left[1-\left(\frac{m_{l}}{m_{\pi}}\right)^{2}\right]^{2}.
\end{equation}
With an appropriate phase convention on the particle states, the 
decay constants can be taken to be real positive numbers which 
have the values $f_{\pi}=(0.1307 \pm 0.00046) $ GeV and 
$f_{K}=(0.1598 \pm 0.00184)$ GeV. Insofar as parity is a good quantum number,
the vector current matrix elements for $\pi$ and $K$ are zero.
By contrast, for the vector mesons $\rho$, $\omega$, $K^{*}$ and $\phi$, 
the axial vector matrix elements are zero. The vector meson decay constants
are defined by
\begin{eqnarray}
\label{fact16}
\langle \rho^{-} | \bar{d}_{\alpha} \gamma^{\mu} u_{\alpha} | 0 \rangle
& = & f_{\rho} m_{\rho} \epsilon ^{\mu}, \qquad
\langle K^{*\;-}| \bar{s}_{\alpha} \gamma^{\mu} u_{\alpha} | 0 \rangle
 =  f_{K^{*}}m_{K^{*}}\epsilon^{\mu}, \nonumber  \\
\langle \phi |\bar{s}_{\alpha} \gamma^{\mu} s_{\alpha} | 0 \rangle
& = & f_{\phi}m_{\phi} \epsilon^{\mu}, \nonumber  \\
\langle \omega | \bar{u}_{\alpha} \gamma^{\mu} u_{\alpha} | 0 \rangle
& = & \langle \omega | \bar{d}_{\alpha} \gamma^{\mu} d_{\alpha} | 0 \rangle
= \frac{1}{\sqrt{2}} f_{\omega} m_{\omega} \epsilon^{\mu}
\end{eqnarray}
where $\epsilon^{\mu}$ is the meson polarization vector. Isospin symmetry
determines the other matrix elements, for example
\begin{equation}
\label{fact17}
\langle \rho^{0} | \bar{u}_{\alpha} \gamma^{\mu} u_{\alpha} | 0 \rangle
= \frac{1}{\sqrt{2}} f_{\rho} m_{\rho} \epsilon^{\mu}.
\end{equation}
The magnitudes of some vector decay constants can be 
inferred from measurements of $\tau$ 
lepton decay, for example
\begin{equation}
\label{fact18}
\fl
\Gamma(\tau^{-} \rightarrow K^{*\;-} \nu_{\tau})  =  
\frac{G_{F}^{2}}{16\pi}\; f_{K^{*}}^{2} \;m_{\tau}^{3} |V_{us}|^{2}
\left[1-\left(\frac{m_{K^{*}}}{m_{\tau}}\right)^{2}\right]^{2}
\left[1+2\left(\frac{M_{K^{*}}}{m_{\tau}}\right)^{2}\right]
\end{equation}
and others from the meson decay rates into $e^{+}e^{-}$ pairs:
\begin{eqnarray}
\label{fact19}
\Gamma(\rho^{0} \rightarrow e^{+}e^{-}) = \frac{2 \pi}{3} \alpha^{2}
\frac{f_{\rho}^{2}}{m_{\rho}}, \nonumber  \\
\Gamma(\omega \rightarrow e^{+}e^{-}) = \frac{1}{9} \frac{2 \pi}{3} 
\alpha^{2}\frac{f_{\omega}^{2}}{m_{\omega}}, \nonumber  \\
\Gamma(\phi \rightarrow e^{+}e^{-}) = \frac{2}{9} \frac{2 \pi}{3} 
\alpha^{2}\frac{f_{\phi}^{2}}{m_{\phi}}.
\end{eqnarray}
With a phase convention that makes the decay constants real and positive,
these yield the well determined values $f_{\rho}=(0.216 \pm 0.005)$ GeV, 
$f_{\omega}=(0.194 \pm 0.004)$ GeV, $f_{\phi}=(0.233 \pm 0.004)$ GeV and
$f_{K^{*}}=(0.216 \pm 0.010)$ GeV. 

We now consider the $B$ transition form factors. For pseudoscalar mesons, 
these are usually expressed in terms of two form factors $F_{0}(t)$
and $F_{1}(t)$:
\begin{equation}
\label{fact20}
\fl
\langle P |\bar{q}_{\alpha}\gamma^{\mu} b_{\alpha}| B \rangle
  =  \left[(p_{B}+p_{P})^{\mu}-\frac{(m_{B}^{2}-m_{P}^{2})}{t} 
 q^{\mu}\right]  F_{1}(t)  
 +\frac{(m_{B}^{2}-m_{P}^{2})}{t} q^{\mu}\; F_{0}(t).
\end{equation}
Here $q^{\mu} \equiv (p_{B}-p_{P})^{\mu}$ and $t \equiv q^{2}$. The 
axial vector matrix elements are all zero. Both $F_{0}$ and $F_{1}$ are
analytical functions of $t$ with no singularity at $t=0$. As there is no
singularity in the matrix element, we have the constraint 
$F_{0}(0) = F_{1}(0)$. The nearest singularity is for $t$ real and greater
than $m_{B}^{2}$, distant from $t=0$. Both $F_{0}$ and $F_{1}$ are often
taken as simple pole or dipole dominant. For example, from lattice QCD,
\begin{equation}
\label{fact21}
F_{1}(t) = \frac{F(0)}{(1-t/m^{2})^{2}}, \qquad
F_{0}(t) = \frac{F(0)}{(1-t/m^{2})}
\end{equation}
with $m^{2} > m_{B}^{2}$. A virtue of the rather clumsy parameterization in
(\ref{fact20})  is that when a contraction is taken with the matrix 
element for the second decay meson then only $F_{0}$ contributes if the
second meson is also a pseudoscalar and only $F_{1}$ contributes if the
second meson is a vector. For example,
\begin{eqnarray}
\label{fact22}
\langle \pi^{-} | \bar{d}_{\alpha} \gamma_{5} \gamma^{\mu} u_{\alpha} 
| 0 \rangle \langle P | \bar{q}_{\beta} \gamma_{\mu} b_{\beta} | B \rangle
& = & - f_{\pi} F_{0}(m_{\pi}^{2}) (m_{B}^{2}-m_{P}^{2}), \\
\label{fact23}
\langle \rho^{-} | \bar{d}_{\alpha} \gamma^{\mu} u_{\alpha} |0 \rangle
\langle P | \bar{q}_{\beta} \gamma_{\mu} b_{\beta}| B \rangle
& = &  2 |p_{\rho}| m_{B} f_{\rho} F_{1}(m_{\rho}^{2})
\end{eqnarray}
where the momentum of the $\rho $ is given by
\begin{equation}
\label{fact24}
|p_{\rho}| = \left[\left(\frac{m_{B}^{2}+m_{\rho}^{2}-m_{P}^{2}}
{2m_{B}}\right)^{2}- m_{\rho}^{2} \right]^{1/2}.
\end{equation}
We have used $\epsilon \cdot p_{B}=|p_{\rho}|m_{B}/m_{\rho}$ in the
$B$ rest frame and taken the $\rho $ meson to be moving along the $z$ axis
with zero helicity so that
\begin{equation}
\label{fact25a}
m_{\rho}\epsilon^{\mu}_{\rho} = (p_{\rho},0,0,E_{\rho}).
\end{equation}

If terms in $(m_{\pi}/m_{b})^{2}=0.0007,\;(m_{\rho}/m_{B})^{2}=0.0212$, e.t.c.
are neglected, as is appropriate for the heavy quark limit, then, because
of the analytic structure of the form factors and the constraint 
$F_{0}(0)=F_{1}(0)$, we can write
\begin{equation}
\label{fact25}
\langle \pi^{-} | (\bar{d}_{\alpha}\;u_{\alpha})_{{\rm V-A}}| 0 \rangle
\langle P | (\bar{q}_{\beta}\;b_{\beta})_{{\rm V-A}} | B \rangle
= - f_{\pi} m_{B}^{2} F_{1}(0)
\end{equation}
and
\begin{equation}
\label{fact26}
\langle \rho^{-} | (\bar{d}_{\alpha}\;u_{\alpha})_{{\rm V-A}}| 0 \rangle
\langle P | (\bar{q}_{\beta}\;b_{\beta})_{{\rm V-A}} | B \rangle
= f_{\rho} m_{B}^{2} F_{1}(0).
\end{equation}
Apart from the well determined $f_{\pi}$ and $f_{\rho}$, in the heavy 
quark limit the $B \rightarrow PP$ and the $B \rightarrow PV$ transitions
through the term (\ref{fact26}) 
are characterized by a single parameter. The use of two parameters is 
not consistent with the heavy quark limit.

The magnitudes of some form factors are measurable in principle 
through semileptonic decays
such as $\bar{B}^{0} \rightarrow \pi^{+} + l^{-} + \bar{\nu}_{e}$,
corresponding to $P=\pi^{+}, \;\bar{q}=\bar{u}$, for which, neglecting terms
proportional to the lepton mass,
\begin{equation}
\label{fact27}
\frac{\rmd \Gamma}{\rmd t} = \frac{G_{F}^{2}}{24\pi^{3}}|V_{ub}|^{2} 
|p_{\pi}| |F_{1}(t)|^{2}.
\end{equation}
However, these relations have only been used to estimate the CKM matrix 
element, the form factors have been taken from theory.   For example,
lattice QCD has been used to estimate $\langle \pi^{+}|\bar{u}_{\alpha}
\gamma^{\mu} b_{\alpha} | \bar{B}^{0} \rangle $ at large values of $t$ 
where the pion is moving slowly. Various phenomenological forms, such as 
(\ref{fact21}) which interpolate quite well through the calculated values,
are then used to extrapolate to the small $t$ region. Table 2 shows 
values for $F_{\pi}(0)$ and $F_{K}(0)$ from lattice QCD and other more
phenomenological estimates. Other transition form factors are related by
isospin symmetry, for example
\begin{equation}
\label{fact28}
\langle \pi^{0} | \bar{u}_{\alpha} \gamma^{\mu} b_{\alpha}| B^{-} \rangle
= \frac{1}{\sqrt{2}} \langle \pi^{+} | \bar{u}_{\alpha} \gamma^{\mu}
b_{\alpha} | \bar{B}^{0} \rangle.
\end{equation}

The transition matrix elements to vector mesons are usually expressed 
through four form factors:
\begin{eqnarray}
\label{fact29}
\fl
\langle V |\bar{q}_{\alpha} \gamma_{\mu} (1-\gamma_{5}) b_{\alpha}|B
\rangle  = &  2\rmi \varepsilon_{\mu \nu \rho \sigma}\;\epsilon^{\nu}
p^{\rho}_{V}p^{\sigma}_{B}\;\frac{V(t)}{m_{B}+m_{V}} \nonumber  \\
& - \epsilon^{\mu} (m_{B}+m_{V})A_{1}(t) 
 +(p_{B}+p_{V})^{\mu} \; \epsilon \cdot q \;
\frac{A_{2}(t)}{m_{B}+m_{V}} \nonumber  \\
& + q^{\mu}\; \epsilon \cdot q \;\frac{2 m_{V}}{t} [A_{3}(t)- A_{0}(t)]
\end{eqnarray}
where
\begin{equation}
\label{fact30}
A_{3}(t)= \left(\frac{m_{B}+m_{V}}{2m_{V}}\right) A_{1}(t) - 
\left(\frac{m_{B}-m_{V}}{2m_{V}}\right) A_{2}(t).
\end{equation}
With an appropriate phase convention all the form factors can be taken
to be real.
Again the form factors are dimensionless analytic functions of $t$ with
the nearest singularity at $t$ real and greater than $m_{B}^{2}$. Also,
the analytic structure demands that $A_{3}(0)=A_{0}(0)$. 
In the semileptonic decays the matrix
elements for the vector mesons to have helicity $+1,\;-1$ or $0$, denoted
by $H_{+}(t)$, $H_{-}(t)$ and $H_{0}(t)$ respectively, are given by
\cite{Behrens2000}
\begin{eqnarray}
\label{fact31}
H_{\pm}(t) & = & (m_{B}+m_{V})A_{1}(t) \mp 
\frac{2 m_{B}|p_{V}|}{m_{B}+m_{V}}\;V(t), \nonumber  \\
H_{0}(t) & = & \frac{1}{2 m_{V}\sqrt{t}}
\left[(m_{B}^{2}-m_{V}^{2}-t)(m_{B}+m_{V})A_{1}(t) 
\frac{4 m_{B}^{2}|p_{V}|^{2}A_{2}(t)}{m_{B}+m_{V}}\right]
\end{eqnarray}
where the vector meson momentum $|p_{V}|$ in the $B$ rest frame is
\begin{equation}
\label{fact32}
|p_{V}| = \left[\left(\frac{m_{B}^{2}+m_{V}^{2}-t}{2 m_{B}}\right)^{2} 
- m_{V}^{2}\right]^{1/2}.
\end{equation}
Note that $|p_{V}|,\;H_{\pm}(t)$ and $\sqrt{t}H_{0}(t)$ are analytic
functions of $t$ with singularities distant from $t=0$. For small $t$ it
appears to be the case \cite{Behrens2000} 
that $A_{1},\;A_{2}$ and $V$ are of similar
magnitude. Hence, for $t \leq m_{B}^{2}$, and barring excessive 
cancellation, it can be anticipated that $\sqrt{t}H_{0}(t)/m_{V}$ will
be larger than $H_{\pm}(t)$ by a factor of $(m_{B}/m_{V})^{2}$.

The squares of some helicity matrix elements can in principle be measured
from the semileptonic decays, for example, 
$\bar{B}^{0} \rightarrow \rho^{+}+l^{-}+\nu $. The decay rate
for the lepton pair to have an invariant mass squared of $t$  is
\begin{equation}
\label{fact33}
\fl
\frac{\rmd \Gamma}{\rmd t}  =  \frac{G_{F}^{2}}{256 \pi^{3}}
\frac{t |p_{V}||V_{qb}|^{2}}{m_{B}^{2}} \;[H_{+}^{2}(1 - \cos \theta )^{2} 
+H_{-}^{2}(1 + \cos \theta )^{2} +2H_{0}^{2} \sin ^{2} \theta ]
\end{equation}
where $\theta $ is the angle between the charged lepton velocity and the 
recoil momentum of the vector boson $V=\rho^{+}$ 
in the lepton pair rest frame.

As with pseudoscalar transitions, these relations (\ref{fact33}) have only
been used to estimate the CKM matrix element $|V_{ub}|$, the form factors
have been taken from theory such as lattice QCD. The CLEO collaboration 
\cite{Behrens2000} have made such an analysis with several theoretical models.
All models but one show a substantial dominance of $H_{0}(t)$ at small
$t$. Table 3 provides various theoretical estimates for the form factors
and helicity matrix elements associated with the  
$\bar{B}^{0} \rightarrow \rho^{+} $ and 
$B^{-} \rightarrow K^{*-}$  vector decays.

Contraction of the transition matrix element (\ref{fact29}) with that for
a pseudoscalar factorization partner gives
\begin{equation}
\label{fact34}
\langle P |J^{\mu} | 0 \rangle \langle V,\lambda =0|J_{\mu}| B \rangle
= 2 m_{B} |p_{V}| f_{P} A_{0}(m_{P}^{2}).
\end{equation}
Here $\lambda =0$ indicates that the vector meson must have zero helicity.
For a vector factorization partner there are three possibilities. Since
the $B$ meson has no spin, both vector particles must have the same
helicity so that
\begin{equation}
\label{fact35}
\langle V_{2}, \lambda |J^{\mu} | 0 \rangle 
\langle V_{1}, \lambda |J_{\mu} | B \rangle 
= f_{V_{2}} m_{V_{2}} H_{\lambda} (m_{V_{2}}^{2})
\end{equation}
where $\lambda = \pm 1, 0$. In the heavy meson limit both mesons should 
have zero helicity. Although some cancellation between form factors can
be anticipated, Table 3 suggests that the helicity zero states will 
dominate the decay rates. Also
\begin{equation}
\label{fact36}
A_{3}(t)  =  \left(\frac{1}{m_{B}^{2}-m_{V}^{2}-t}\right) 
  \left[\sqrt{t}H_{0}(t)-\frac{m_{B}^{2}+3m_{V}^{2}-t}
 {2 m_{V}(m_{B}+m_{V})}\;A_{2}(t)\right]
\end{equation}
so that, from (\ref{fact31}), the constraint $A_{3}(0)=A_{0}(0)$ and the
fact that the singularities are distant from the small $t$ region, we
can write, for example, in the heavy meson limit where terms in 
$(m_{V}/m_{B})^{2}$ are neglected,
\begin{equation}
\label{fact37}
\langle \pi^{-} | (\bar{d}_{\alpha}\;u_{\alpha})_{{\rm V-A}}| 0 \rangle
\langle V,0 | (\bar{u}_{\beta}\;b_{\beta})_{{\rm V-A}} | B \rangle
= f_{\pi} m_{B}^{2} A_{0}(0)
\end{equation}
and
\begin{equation}
\label{fact38}
\langle \rho^{-} | (\bar{d}_{\alpha}\;u_{\alpha})_{{\rm V-A}}| 0 \rangle
\langle V,0 | (\bar{u}_{\beta}\;b_{\beta})_{{\rm V-A}} | B \rangle
=  f_{\rho} m_{B}^{2} A_{0}(0).
\end{equation}
Thus the four soft QCD parameters characterizing decays into vector mesons
reduce to just one in this limit.

Based upon the simplifications discussed above, we show in tables 4 to 7 
our expressions for the factorization matrix elements
\begin{equation}
\label{fact39}
\tilde{Q}_{i} (h_{1},h_{2}) \equiv m_{B}^{-2}[Q_{i}(h_{1},h_{2})
+Q_{i}(h_{2},h_{1})]
\end{equation}
for a wide range of $B$ decays. In these tables we have used the notation
$F_{\pi}=F_{1}^{\pi}(0), \; A_{\rho} = A_{0}^{\rho}(0)$, etc.
and include chiral enhancement factors
$R^{\pi}_{\chi}=2 m_{\pi}^{2}/[m_{b}(m_{u}+m_{d})]$ and 
$R^{K}_{\chi}=2 m_{K}^{2}/[m_{b}(m_{u}+m_{s})]$ 
for the $a_{6}$ contributions.

\section{Sign conventions for decay constants and form factors}

It is clear in the literature (see, for example, references 
\cite{AliGreub98,Alietal98,Behrens2000}) that different authors use 
different phase conventions for the particle states in defining the
current matrix elements. Changes of convention should only multiply
the combination $\sum_{i=1, \ldots ,6,9}p_{i}\tilde{Q}_{i}$ of the
matrix elements $\tilde{Q}_{i}$ occurring in (\ref{fact40}) by a common
phase factor, thus leaving the branching ratios unaltered. However, an
inconsistent convention, such as defining $f_{\pi}$ through
$\langle \pi^{+}(p)|\bar{u}_{\alpha}\gamma_{5}\gamma^{\mu}d_{\alpha}|0\rangle
=f_{\pi}p^{\mu}_{\pi}$ but insisting that $f_{\pi}$ is positive, will result
in different branching ratios. If the experimental program, outlined in 
this paper, were to be carried through, the relative signs of these soft QCD
parameters would not be determined and at this preliminary stage of $B$ 
decay data analysis a theory must be consulted to determine these signs.
We outline here a simple quark model that illustrates this procedure.

Consider the current operator $j^{\mu}(0)=\bar{u}\gamma^{\mu}(1-\gamma_{5})d$
where the $u$ and $d$ quark fields are evaluated at $x^{\mu}=0$. To construct
the matrix elements $\langle \pi^{+}|j^{\mu}|0\rangle $ and
$\langle \rho^{+}|j^{\mu}|0\rangle $ we take $|0\rangle$ to be the state
with no quarks or antiquarks and $|\pi^{+} \rangle $ and $|\rho^{+} \rangle $ 
to be a $u\bar{d} $ pair at rest with a bound state $S$ wave function 
$\phi(r)$ for their relative distance $r$. The $\pi^{+}$ and $\rho^{+}$ 
spin states are
\begin{equation}
\label{app1}
| \pi^{+}\rangle _{{\rm spin}} =\case{1}{\sqrt{2}} 
[|u,\case{1}{2}\rangle |\bar{d},-\case{1}{2} \rangle
- |u,-\case{1}{2}\rangle |\bar{d}, \case{1}{2} \rangle ]
\end{equation}
and
\begin{equation}
\label{app2}
| \rho^{+}\rangle _{{\rm spin}} = \case{1}{\sqrt{2}}
[|u,\case{1}{2}\rangle |\bar{d},-\case{1}{2} \rangle
+ |u,-\case{1}{2}\rangle |\bar{d}, \case{1}{2} \rangle ].
\end{equation}
We then find, up to an overall positive dimensionless factor,
\begin{equation}
\label{app3}
\langle \pi^{+} | j^{\mu} |0\rangle \propto \phi^{*}_{\pi}(0)[-1,0,0,0]
\end{equation}
and
\begin{equation}
\label{app4}
\langle \rho^{+} | j^{\mu} |0\rangle \propto \phi^{*}_{\rho}(0)[0,0,0,1].
\end{equation}
A comparison with (\ref{fact13}), (\ref{fact16}) and (\ref{fact25a}) 
then gives
\begin{equation}
\label{app5}
f_{\pi} \propto \phi^{*}_{\pi}(0)/\sqrt{m_{\pi}}
\end{equation}
and
\begin{equation}
\label{app6}
f_{\rho} \propto \phi^{*}_{\rho}(0)/\sqrt{m_{\rho}}.
\end{equation}

Similarly, we construct the matrix elements 
$\langle\pi^{+} | \bar{u}\gamma^{\mu}(1-\gamma_{5})b|\bar{B}^{0}\rangle $
and
$\langle\rho^{+} | \bar{u}\gamma^{\mu}(1-\gamma_{5})b|\bar{B}^{0}\rangle $
by assuming that the $\bar{B}^{0}$ is at rest with a $b\bar{d}$ $S$ wave
function $\phi_{B}(r)$ and spin state
\begin{equation}
\label{app7}
| \bar{B}^{0} \rangle _{{\rm spin}} = \case{1}{\sqrt{2}}
[|b,\case{1}{2}\rangle |\bar{d},-\case{1}{2} \rangle
- |b,-\case{1}{2}\rangle |\bar{d}, \case{1}{2} \rangle ].
\end{equation}
For the $\pi^{+}$ and $\rho^{+}$ again at rest we find, up to an overall
positive factor,
\begin{equation}
\label{app8}
\langle \pi^{+} | j^{\mu} |\bar{B}^{0} \rangle 
\propto \int \phi^{*}_{\pi}(r) \phi_{B}(r)r^{2}\;\rmd r\;[1,0,0,0]
\end{equation}
and
\begin{equation}
\label{app9}
\langle \rho^{+} | j^{\mu} |\bar{B}^{0} \rangle 
\propto \int \phi^{*}_{\rho}(r) \phi_{B}(r)r^{2}\;\rmd r\;[0,0,0,-1]
\end{equation}
so that, from (\ref{fact20}), (\ref{fact29}) and (\ref{fact25a}), we infer
\begin{equation}
\label{app10}
F_{0}((m_{B}-m_{\pi})^{2}) \propto \int \phi^{*}_{\pi} (r) \phi_{B}(r)
r^{2}\;\rmd r
\end{equation}
and
\begin{equation}
\label{app11}
A_{1}((m_{B}-m_{\rho})^{2}) \propto \int \phi^{*}_{\rho} (r) \phi_{B}(r)
r^{2}\;\rmd r.
\end{equation}
Noting that the form factors presented in the literature do not change 
sign on extrapolation to $t=0$ then we observe that taking the wave 
functions to be all real and positive is in accord with our sign
conventions for the four parameters $f_{\pi}, f_{\rho}, F_{0}$ and
$A_{1}$.

\section{Confronting the model with data}

The branching ratios for two-body $B$ decays are given, in the heavy mass
limit, by
\begin{equation}
\label{fact40}
{\rm Br}(B \rightarrow h_{1}h_{2})  =  S \frac{G_{F}^{2}m_{B}^{3}}
{32 \pi \Gamma_{{\rm total}}}\;
\left|\sum_{i=1,\ldots ,6,9} p_{i}\; \tilde{Q}_{i}(h_{1},h_{2})\right|^{2}
\end{equation}
where $S=1$ unless the two bodies are identical, in which case the 
angular phase space is halved and $S=1/2$.
We have attempted to fit the theoretical expressions for branching ratios
with available data. Measured branching ratios for nineteen channels are
shown in table 8, along with references to where the data can be found.
For many channels there are measurements presented by more than one of
the three groups CLEO, BaBar and Belle. We have not attempted to combine
the results, for each channel the table shows the measured branching ratio
with the smallest quoted errors. We take the measured branching ratios
to be the mean of the $B$ and $\bar{B}$ decays:
\begin{equation}
\label{fact51}
{\rm Br}({\rm exp})=\frac{1}{2} \left[{\rm Br}(B \rightarrow h_{1}h_{2})
+ {\rm Br} (\bar{B} \rightarrow \bar{h}_{1} \bar{h}_{2})\right].
\end{equation}
We ignore here the CP asymmetries
\begin{equation}
\label{fact41}
A_{{\rm CP}}= \frac{{\rm Br}(B \rightarrow h_{1}h_{2})
- {\rm Br}(\bar{B} \rightarrow \bar{h}_{1}\bar{h}_{2})}
{{\rm Br}(B \rightarrow h_{1}h_{2})+{\rm Br}(\bar{B} \rightarrow
\bar{h}_{1}\bar{h}_{2})}
\end{equation}
which are, so far, all consistent with being zero.

For convenience we assign to each channel $(h_{1}h_{2})$ a number $\alpha $.
The statistical and systematic errors have been combined into a single error
$\sigma_{\alpha}$. The systematic errors in particular can be expected
to be correlated. Here we ignore all correlations and form a
$\chi ^{2}$ function
\begin{equation}
\label{fact52}
\chi ^{2}(P_{i}) = \sum_{\alpha} \left[ \left|{\rm Br}_{\alpha} (P_{i})-
{\rm Br}_{\alpha} ({\rm exp})\right|/\sigma_{\alpha}\right] ^{2}
+ {\rm additional\;constraints}.
\end{equation}
${\rm Br}_{\alpha}(P_{i})$ are the theoretical branching ratios given by
(\ref{fact40}) in terms of nine parameters $P_{i}, i=1,\ldots ,9 $ which
we take to be the three Wolfenstein CKM parameters
$\{A, \bar{\rho}, \bar{\eta}\}$ and the
six soft QCD parameters $\{R^{K}_{\chi},F_{\pi}, F_{K},
A_{\rho},A_{\omega},A_{K^{*}}\}$. The small contribution of the
$b \rightarrow u $
penguin makes the branching ratios insensitive to $R^{\pi}_{\chi}$,which we
hold fixed at $R^{\pi}_{\chi} = 0.97$. The well known decay parameters
$\{f_{\pi},f_{K},f_{\rho},f_{\omega},f_{\phi},f_{K^{*}}\}$ are held at their
mean values and the Wolfenstein CKM $\lambda $ parameter is taken to be
$\lambda = 0.2205$. Additional constraints were added to the $\chi ^{2}$
to take into account experimental and theoretical results lying outside
the data on $B$ decay branching ratios.
For example, we took the Wolfenstein parameter $A$ to be close to
$0.802$ as has been inferred from many weak decay measurements.
We search for a minimum of $\chi ^{2}$ as a function of the $P_{i}$ where the
 minimum is close to the expected values of the soft QCD parameters given in 
tables 2 and 3. We use the MINUIT~\cite{MINUIT} program to minimise the 
$\chi ^{2}$.

The theoretical branching ratios and contributions of the individual channels
to $\chi^{2}$ based on these best fit values are given in table 8.
The theoretical values shown use the $a_{i}$ coefficients listed in table 1
as model 2. These are  the process-independent generalized factorization
$a_{i}$ coefficients computed for the renormalization scale $\mu =m_{b}/2$.
Also in the fit we take the electroweak penguin contribution to be
$a_{9}=-0.0094-0.0002i$ \cite{Alietal98}.
Table 8 shows results with the systematic and statistical errors added in 
quadrature:$\sigma ^{2} = \sigma^{2}_{{\rm stat}}+\sigma^{2}_{{\rm syst}}$.
We have also done the computation using simple addition of errors
$\sigma = \sigma_{{\rm stat}} + \sigma_{{\rm syst}}$. Both procedures are
ad hoc, addition in quadrature reduces the influence of the systematic errors
which are in general the smallest.
The values of the best fit parameters $P_{2,i}$ are shown in table 9 together
with our estimates of the two standard deviation errors.  These errors are of 
course highly correlated. A plot of the error matrix ellipse for the
Wolfenstein
parameters $\bar{\rho}$ and $\bar{\eta}$ is shown in figure 1.
The results in table 9 for the best
fit values of the various form factors lie within the spread of theoretical 
estimates for these form factors (see tables 2 and 3).

\section{Discussion and conclusions}

We have investigated two-body charmless hadronic $B$ decays within the
so called QCD factorization model, making use of simplifications which
arise from working in the heavy quark limit. This is particularly 
evident for the $B \rightarrow VV$ processes which we argue to be 
predominantly of zero helicity. Consequently a wide range of decays
can be expressed in terms of a relatively small number of soft QCD
parameters, thus providing a theoretical framework which should be
adequate to confront data for some time to come.

Different factorization models merely modify the $a_{i}$ coefficients 
which premultiply the various combinations of soft QCD parameters, 
thus allowing ready comparisons between these models.

If, in $B$ decays to two vector mesons, there is a significant contribution
from the $\lambda = \pm $ helicity states, it should be apparent in the
Dalitz type plots for the final decay products. Table 3 suggests that 
the negative helicity state might be important for $\bar{B}^{0}$ and
$B^{-}$ decays, and the positive helicity state for $B^{0}$ and $B^{+}$
decays. Since each helicity contributes incoherently to the branching
ratio, each helicity can be considered as a separate channel. The 
additional helicity channels can be included at the cost of extra 
soft QCD parameters.
The only vector channels in table 8 are $B \rightarrow K^{*} \phi $ and in
these channels there is some evidence \cite{Ba0105} for contributions
from non-zero helicities. It can be seen from tables 1, 6 and 7 that
the zero helicity amplitudes (the only ones included) are proportional to
$A_{K^{*}}$, a parameter which contributes significantly only to the
$K^{*} \phi $ channels. Splitting the decay rates into the individual helicity
channels will hardly effect the fit, it will only modify the estimate of
$A_{K^{*}}$  given in table 9 and introduce more soft QCD parameters for
the other helicities.

To economize in the number of soft QCD parameters we have not included
decay channels involving $\eta$ and $\eta^{\prime}$ mesons. These 
amplitudes involve the mixing angle between the $(u\bar{u}+d\bar{d})$ and 
$s\bar{s}$ combinations. Also, in principle, there is mixing with $c\bar{c}$
which, though small, could make a significant contribution to decay modes
through the enhanced quark decay modes $b \rightarrow cq\bar{c}$.

From table 8 it seems that of the nineteen channels included in the present
analysis only the $PV$ channel $\pi^{-}K^{*0}$, and to a lesser extent 
$\pi^{-}K^{*+}$ and $\omega K^{0}$, give a large contribution to the overall
$\chi^{2}$. The $\omega K^{0}$ channel has its largest theoretical contribution
from the $b \rightarrow s $ penguin and, in particular, from the $a_{4}$ and 
$a_{6}$ terms. The theoretical branching ratio is small because of the 
cancellation, evident in tables 1 and 6, between these terms.  It is difficult 
for the theory to explain a branching ratio greater than $2 \times 10^{-6}$. 

The $\pi^{-}K^{*0}$ channel is well measured, Belle gives
a large branching
ratio consistent with the BaBar result $(15.5 \pm 3.4 \pm 1.8)\times 10^{-6}$
shown in table 8, whereas the CLEO value is lower.
It is interesting to compare this channel with $\pi^{-}K^{0}$, also well
measured and with only a marginally larger branching ratio. A fit of the
theoretical ratio
\begin{eqnarray}
\label{fact55}
\frac{{\rm Br}(B \rightarrow \pi^{-}K^{0})}
{{\rm Br}(B \rightarrow \pi^{-}K^{*0})}
& = & \left( \frac{f_{K}}{f_{K^{*}}}\right)
\left(1+R^{K}_{\chi}\frac{a_{6}}{a_{4}}\right)^{2} \nonumber  \\
& \approx & 0.547 (1+ 1.3 R^{K}_{\chi})^{2}
\end{eqnarray}
to the BaBar and Belle data implies $R^{K}_{\chi} \leq 0.5 $. Our fit
does agree well with the CLEO result.

\Tables

\fulltable{Decay amplitude coefficients $a_{i}$ in the matrix 
elements of the effective Hamiltonian (\ref{fact7}) for several theoretical
models. Model 1 is a simple tree plus QCD penguin as described in section 2,
model 2 is generalized factorization with the effective Wilson coefficients  
evaluated at the renormalization scale $\mu = m_{b}/2$ and 
$q^{2}= m_{b}^{2}/4$ \protect \cite{CMW99}, 
and model 3 is QCD factorization at
$\mu = m_{b}/2$ evaluated using the expressions of 
Beneke {\it et al} \protect \cite{Beneke99,Benekeetal2001}.
The imaginary part of each coefficient is in parentheses. }
\label{table1}
\begin{tabular}{@{}llllll}
\br
Model & $a_{1}^u$ & $a_{2}^u$ & $a_{3}^u = a_3^c$ & $a_{4}^u$ & $a_{4}^c$ \\
\mr
1 &  \01.0000 & \0\m0.3333 & \00.0000 & \0$-$0.0492 & \0$-$0.0524 \\
&&&& ($-0.0234i$) & ($-0.0059i$)  \\
2 & \01.0523 & \0\m0.0333 & \00.0062  & \0$-$0.0397 & \0$-$0.0496 \\
&&&& ($-$0.0148i) & ($-$0.0094i)  \\
3 & \01.0510  & \0\m0.0499 & \00.0050 & \0$-$0.030 & \0$-$0.038\\
& (0.0335i)  & ($-$0.1063i) & (0.0033i)  & ($-$0.019i) & ($-$0.009i)  \\

\mr
& $a_{5}^u = a_5^c$ & $a_{6}^u$ & $a_{6}^c$ &&\\
\mr
1 &  \0\m0.0000 & \0$-$0.0492 & \0$-$0.0524 &&\\
&& ($-0.0234i$)  & ($-0.0059i$)  &&\\
2 & \0$-$0.0064  & \0$-$0.0397 & \0$-$0.0496 &&\\
&& ($-$0.0148i)  & ($-$0.0094i)  &&\\
3 & \0\m0.0050 & \0$-$0.050 & \0$-$0.055 &&\\
& \m(0.0033i)  & ($-$0.017i) & ($-$0.005i) && \\
\br
\end{tabular}
\endfulltable

\begin{table}
\caption{Theoretical form factors for $B \rightarrow \pi$ and 
$B \rightarrow  K$ transitions.}
\label{table2}
\begin{indented}
\item[]\begin{tabular}{@{}lll}
\br
Model & $F_{\pi}(0)$ & $F_{K}(0)$ \\
\mr
Lattice QCD$^{a}$ & 0.27 & \\
Quark model wave functions$^{b}$& 0.33 & \\
Quark model wave functions$^{c}$ & & 0.38  \\
Light cone sum rule$^{d}$ & 0.305 & 0.341 \\
\br
\end{tabular}
\item[]{$^{a}$ Ref. \cite{Debbio98}}
\item[]{$^{b}$ Ref. \cite{Wirbel85}}
\item[]{$^{c}$ Ref. \cite{Bauer87}}
\item[]{$^{d}$ Ref. \cite{Ball98}}
\end{indented}
\end{table}

\fulltable{Theoretical form factors and helicity amplitudes 
for $\bar{B}^{0} \rightarrow \rho^{+} $ and  
$B^{-} \rightarrow K^{*-}$ transitions. It can be seen that, although the
helicity zero channel for $B \rightarrow VV$ decays can be expected to
dominate, this table shows that much cancellation is anticipated in
(\ref{fact31}). The extent to which the helicity zero states dominate
should be apparent in Dalitz type plots for the final decay products.}
\label{table3}
\begin{tabular}{@{}llllllll}
\br
Reference & $A_{1}(0)$ & $A_{2}(0)$ & $V(0)$ & $A_{0}(0)$ &
$|H_{+}(m_{V}^{2})|^{2}$ & $|H_{-}(m_{V}^{2})|^{2}$ &
$|H_{0}(m_{V}^{2})|^{2}$ \\
\mr
$\bar{B}^{0} \rightarrow \rho^{+} $ & & & & & & &  \\
1$^{a}$ & 0.27 & 0.26 & 0.35 & 0.30 & 0.008 & 10.1 & 121 \\
2$^{b}$ & 0.26 & 0.22 & 0.34 & 0.38 & 0.006 & \09.4  & 185  \\
3$^{c}$ & 0.27 & 0.28 & 0.35 & 0.24 & 0.008 & 10.1 & \081  \\
4$^{d}$ & 0.30 & 0.33 & 0.37 & 0.21 & 0.034 & 11.9 & \066  \\
5$^{e}$ & 0.28 & 0.28 & 0.33 & 0.28 & 0.057 &  \09.9 & 107 \\
$B^{-} \rightarrow K^{*-}$ & & & & & & &  \\
2$^{b}$ & 0.34 & 0.28 & 0.46 & 0.49 & 0.02 & 16.4 & 224  \\
4$^{d}$ & 0.36 & 0.40 & 0.45 & 0.26 & 0.10 & 17.1 & \075  \\
6$^{f}$ & 0.37 & 0.40 & 0.47 & 0.30 & 0.30 & 18.3 & \093  \\
7$^{g}$ & 0.33 & 0.33 & 0.37 & 0.32 & 0.22 & 13.0 & 110  \\
\br
\end{tabular}\\
\noindent{$^{a}$ Ref. \cite{Debbio98}}
\noindent{$^{b}$ Ref. \cite{Ball98a}}
\noindent{$^{c}$ Ref. \cite{Ball97}}
\noindent{$^{d}$ Ref. \cite{Aliev97}}
\noindent{$^{e}$ Ref. \cite{Wirbel85}}
\noindent{$^{f}$ Ref. \cite{Colangelo96}}
\noindent{$^{g}$ Ref. \cite{Bauer87}}
\endfulltable

\begin{table}
\caption{Factorization matrix elements $\tilde{Q}_{i}(h_{1},h_{2})$ for
$\bar{B}^{0} \rightarrow h_{1}h_{2}$ decays arising from 
$b \rightarrow d q \bar{q}$.}
\label{table4}
\lineup
\begin{tabular}{@{}llllllll}
\br
Decay$^{a}$  & $\tilde{Q}_{1}$ & $\tilde{Q}_{2}$ & $\tilde{Q}_{3}$ &
$\tilde{Q}_{4}$ & $\tilde{Q}_{5}$ & $\tilde{Q}_{6}$ &
$\tilde{Q}_{9}$ \\
\mr
$\pi^{+}\pi^{-}$ & $-F_{\pi}f_{\pi}$ & 0 & 0 & $-F_{\pi}f_{\pi}$ & 0 & 
$-R_{\chi}^{\pi}F_{\pi}f_{\pi}$ & 0 \\
$\pi^{0}\pi^{0}$ & 0 & $F_{\pi}f_{\pi}$ & 0 & $-F_{\pi}f_{\pi}$ & 0 &
$-R_{\chi}^{\pi}F_{\pi}f_{\pi}$ & $\frac{3F_{\pi}f_{\pi}}{2}$ \\
$\rho^{+}\pi^{-}$ & $A_{\rho}f_{\pi}$ & 0 & 0 & $A_{\rho}f_{\pi}$ & 0 &
$-R_{\chi}^{\pi}A_{\rho}f_{\pi}$ & 0  \\
$\rho^{0}\pi^{0}$ & 0 & $-\frac{A_{\rho}f_{\pi}+F_{\pi}f_{\rho}}{2}$ & 
0 & $ \frac{A_{\rho}f_{\pi}+F_{\pi}f_{\rho}}{2}$ & 0 & 
$-R_{\chi}^{\pi}\frac{A_{\rho}f_{\pi}}{2}$ &
$ -\frac{3(A_{\rho}f_{\pi}+ F_{\pi}f_{\rho})}{4}$  \\
$\pi^{+}\rho^{-}$ & $F_{\pi}f_{\rho}$ & 0 & 0 & $F_{\pi}f_{\rho}$ & 0 & 
0 & 0  \\
$\omega \pi^{0}$ & 0 & $\frac{A_{\omega}f_{\pi}-F_{\pi}f_{\omega}}{2} $ & 
$-F_{\pi}f_{\omega}$ & $-\frac{A_{\omega}f_{\pi}+F_{\pi}f_{\omega}}{2}$ &
$-F_{\pi}f_{\omega}$ & $ R_{\chi}^{\pi}\frac{A_{\omega}f_{\pi}}{2}$ 
& $\frac{3 A_{\omega}f_{\pi}-F_{\pi}f_{\omega}}{4} $ \\
$\rho^{+}\rho^{-}$ & $A_{\rho}f_{\rho}$ & 0 & 0 & $A_{\rho}f_{\rho}$ & 
0 & 0 & 0  \\
$\rho^{0}\rho^{0}$ & 0 & $-A_{\rho}f_{\rho}$ & 0 & $A_{\rho}f_{\rho}$ & 
0 & 0 & $-\frac{3A_{\rho}f_{\rho}}{2}$  \\
$\omega \rho^{0}$ & 0 & $\frac{A_{\omega}f_{\rho}-A_{\rho}f_{\omega}}{2}$
& $-A_{\rho}f_{\omega}$ & $-\frac{A_{\omega}f_{\rho}+A_{\rho}f_{\omega}}{2}$
& $-A_{\rho}f_{\omega}$ & 0 &
$\frac{3 A_{\omega}f_{\rho}-A_{\rho}f_{\omega}}{4}$  \\
$\omega \omega$ & 0 & $A_{\omega}f_{\omega}$ & $2A_{\omega}f_{\omega}$ &
$A_{\omega}f_{\omega}$ & $2A_{\omega}f_{\omega}$ & 0 &
$\frac{A_{\omega}f_{\omega}}{2}$  \\
$\bar{K}^{0} K^{0}$ & 0 & 0 & 0 & $-F_{K}f_{K}$ & 0 
& $-R_{\chi}^{K}F_{K}f_{K}$ & 0  \\
$\bar{K}^{*0}K^{0}$ & 0 & 0 & 0 & $A_{K^{*}}f_{K}$ & 0 
& $-R_{\chi}^{K}A_{K^{*}}f_{K}$
& 0  \\
$\bar{K}^{0}K^{*0}$ & 0 & 0 & 0 & $F_{K}f_{K^{*}}$ & 0 & 0 & 
0  \\
$\bar{K}^{*0}K^{*0}$ & 0 & 0 & 0 & $A_{K^{*}}f_{K^{*}}$ & 0 & 0 & 
0  \\
$\phi \pi^{0}$ & 0 & 0 & 
$\frac{F_{\pi}f_{\phi}}{\sqrt{2}} $ & 0 & 
$\frac{F_{\pi}f_{\phi}}{\sqrt{2}} $ & 0 &
$-\frac{F_{\pi}f_{\phi}}{3\sqrt{2}} $  \\
\br
\end{tabular} \\
\noindent{$^{a}$ The decays to $\phi \rho^{0}$ and $\phi \omega$ are 
obtained from $\phi \pi^{0}$ by the substitutions 
$F_{\pi} \rightarrow A_{\rho}$ and $F_{\pi} \rightarrow -A_{\omega}$ 
respectively.
The decays to $K^{+}K^{-},\;K^{+}K^{*-},\; K^{*+}K^{-}$ and 
$K^{*+}K^{*-}$ receive no contribution from $\tilde{Q}_{1,\ldots, 6}$}
\end{table}

\begin{table}
\caption{Factorization matrix elements $\tilde{Q}_{i}(h_{1},h_{2})$ 
for $B^{-} \rightarrow h_{1}h_{2}$ decays arising from 
$b \rightarrow d q \bar{q}$.} 
\label{table5}
\lineup
\begin{tabular}{@{}llllllll}
\br
Decay$^{a}$  & $\tilde{Q}_{1}$ & $\tilde{Q}_{2}$ & $\tilde{Q}_{3}$ &
$\tilde{Q}_{4}$ & $\tilde{Q}_{5}$ & $\tilde{Q}_{6}$ &
$\tilde{Q}_{9}$ \\
\mr
$\pi^{0}\pi^{-}$ & $-\frac{F_{\pi}f_{\pi}}{\sqrt{2}}$ &  
$-\frac{F_{\pi}f_{\pi}}{\sqrt{2}}$ & 0 & 0 & 0 & 0 &
$-\frac{3F_{\pi}f_{\pi}}{2\sqrt{2}}$  \\
$\rho^{0}\pi^{-}$ &  $\frac{A_{\rho}f_{\pi}}{\sqrt{2}}$ & 
$\frac{F_{\pi}f_{\rho}}{\sqrt{2}}$ & 0 &
$\frac{A_{\rho}f_{\pi}-F_{\pi}f_{\rho}}{\sqrt{2}}$ & 0 & 
$-R_{\chi}^{\pi}\frac{A_{\rho}f_{\pi}}{\sqrt{2}}$ &
$\frac{3F_{\pi}f_{\rho}}{2\sqrt{2}} $  \\
$\omega\pi^{-}$ & $\frac{A_{\omega}f_{\pi}}{\sqrt{2}}$ &  
$\frac{F_{\pi}f_{\omega}}{\sqrt{2}}$ & $\sqrt{2}F_{\pi}f_{\omega}$ & 
$\frac{A_{\omega}f_{\pi}+F_{\pi}f_{\omega}}{\sqrt{2}}$ & 
$ \sqrt{2}F_{\pi}f_{\omega}$ 
& $-R_{\chi}^{\pi}\frac{A_{\omega}f_{\pi}}{\sqrt{2}}$ & 
$\frac{F_{\pi}f_{\omega}}{2\sqrt{2}}$  \\
$\pi^{0}\rho^{-}$ & $\frac{F_{\pi}f_{\rho}}{\sqrt{2}}$ & 
$\frac{A_{\rho}f_{\pi}}{\sqrt{2}}$ & 0 &
$ \frac{F_{\pi}f_{\rho}-A_{\rho}f_{\pi}}{\sqrt{2}}$ & 0 & 
$R_{\chi}^{\pi}\frac{A_{\rho}f_{\pi}}{\sqrt{2}}$ &
$\frac{3A_{\rho}f_{\pi}}{2\sqrt{2}} $  \\
$\rho^{0}\rho^{-}$ & $\frac{A_{\rho}f_{\rho}}{\sqrt{2}}$ & 
$\frac{A_{\rho}f_{\rho}}{\sqrt{2}}$ & 0 & 0 & 0 & 0 &
$\frac{3A_{\rho}f_{\rho}}{2\sqrt{2}} $  \\
$\omega \rho^{-}$ & $\frac{A_{\omega}f_{\rho}}{\sqrt{2}}$  & 
$\frac{A_{\rho}f_{\omega}}{\sqrt{2}} $ & $\sqrt{2}A_{\rho}f_{\omega}$ &
$ \frac{A_{\omega}f_{\rho}+A_{\rho}f_{\omega}}{\sqrt{2}}$ &
$ \sqrt{2}A_{\rho}f_{\omega}$ & 0 &
$\frac{A_{\rho}f_{\omega}}{2\sqrt{2}} $  \\
$K^{-} K^{0}$ & 0 & 0 & 0 & $-F_{K}f_{K}$ & 0 
& $-R_{\chi}^{K}F_{K}f_{K}$ &  0  \\
$K^{*-}K^{0}$ & 0 & 0 & 0 & $A_{K^{*}}f_{K}$ & 0 
& $-R_{\chi}^{K}A_{K^{*}}f_{K}$ & 0  \\
$K^{-}K^{*0}$ & 0 & 0 & 0 & $F_{K}f_{K^{*}}$ & 0 & 0 & 
0 \\
$K^{*-}K^{*0}$ & 0 & 0 & 0 & $A_{K^{*}}f_{K^{*}}$ & 0 & 0 & 
0  \\
$\phi \pi^{-}$ & 0 & 0 & $F_{\pi}f_{\phi}$ & 0 & 
$F_{\pi}f_{\phi} $ & 0 &
$-\frac{F_{\pi}f_{\phi}}{2} $  \\
\br
\end{tabular} \\
\noindent{$^{a}$ The decay to $\phi \rho^{-}$ is obtained from 
that to $\phi \pi^{-}$ by the substitution $F_{\pi} \rightarrow A_{\rho}$.}
\end{table}

\begin{table}
\caption{Factorization matrix elements $\tilde{Q}_{i}(h_{1},h_{2})$ 
for $\bar{B}^{0} \rightarrow h_{1}h_{2}$ decays arising from 
$b \rightarrow s q \bar{q}$.}
\label{table6}
\lineup
\begin{tabular}{@{}llllllll}
\br
Decay  & $\tilde{Q}_{1}$ & $\tilde{Q}_{2}$ & $\tilde{Q}_{3}$ &
$\tilde{Q}_{4}$ & $\tilde{Q}_{5}$ & $\tilde{Q}_{6}$ &
$ \tilde{Q}_{9}$ \\
\mr
$K^{-}\pi^{+}$ & $-F_{\pi}f_{K}$ & 0 & 0 & $-F_{\pi}f_{K}$ & 0 & 
$-R_{\chi}^{K}F_{\pi}f_{K}$ & 0 \\
$K^{*-}\pi^{+}$ &  $F_{\pi}f_{K^{*}}$ & 0 & 0 & $F_{\pi}f_{K^{*}}$ & 
0 & 0 & 0 \\
$K^{-}\rho^{+}$ & $A_{\rho}f_{K}$ & 0 & 0 & $A_{\rho}f_{K}$ & 0 &
$-R_{\chi}^{K}A_{\rho}f_{K}$ & 0  \\
$K^{*-}\rho^{+}$ &  $A_{\rho}f_{K^{*}}$ & 0 & 0 & $ A_{\rho}f_{K^{*}}$ & 
0 & 0  & 0  \\
$\bar{K}^{0} \pi^{0}$ & 0 & $-\frac{F_{K}f_{\pi}}{\sqrt{2}}$  & 0 & 
$\frac{F_{\pi}f_{K}}{\sqrt{2}}$ & 0 
& $R_{\chi}^{K}\frac{F_{\pi}f_{K}}{\sqrt{2}}$ & 
$-\frac{3F_{K}f_{\pi}}{2\sqrt{2}} $  \\
$\bar{K}^{*0}\pi^{0}$ & 0 & $\frac{A_{K^{*}}f_{\pi}}{\sqrt{2}}$ & 0 & 
$-\frac{F_{\pi}f_{K^{*}}}{\sqrt{2}}$ & 0 & 0 &
$\frac{3A_{K^{*}}f_{\pi}}{2\sqrt{2}} $  \\
$\bar{K}^{0}\rho^{0}$ & 0 & $\frac{F_{K}f_{\rho}}{\sqrt{2}}$ & 0 & 
$-\frac{A_{\rho}f_{K}}{\sqrt{2}}$ & 0 
& $R_{\chi}^{K}\frac{A_{\rho}f_{K}}{\sqrt{2}}$ & 
$\frac{3F_{K}f_{\rho}}{2\sqrt{2}} $  \\
$\bar{K}^{*0}\rho^{0}$ & 0 & $\frac{A_{K^{*}}f_{\rho}}{\sqrt{2}}$ & 0 & 
$-\frac{A_{\rho}f_{K^{*}}}{\sqrt{2}}$ & 0 & 0 & 
$\frac{3A_{K^{*}}f_{\rho}}{2\sqrt{2}} $  \\
$\bar{K}^{0}\omega$ & 0 & $\frac{F_{K}f_{\omega}}{\sqrt{2}}$ & 
$\sqrt{2}F_{K}f_{\omega}$  & $\frac{A_{\omega}f_{K}}{\sqrt{2}}$ &
$ \sqrt{2}F_{K}f_{\omega} $ 
& $-R_{\chi}^{K}\frac{A_{\omega}f_{K}}{\sqrt{2}} $ & 
$\frac{F_{K}f_{\omega}}{2\sqrt{2}} $  \\
$\bar{K}^{*0}\omega $ & 0 & $\frac{A_{K^{*}}f_{\omega}}{\sqrt{2}}$ &
$\sqrt{2}A_{K^{*}}f_{\omega}$ & $\frac{A_{\omega}f_{K^{*}}}{\sqrt{2}} $
& $\sqrt{2}A_{K^{*}}f_{\omega}$ & 0 &
$\frac{A_{K^{*}}f_{\omega}}{2\sqrt{2}} $  \\
$\bar{K}^{0}\phi $ & 0 & 0 & $F_{K}f_{\phi}$ & $F_{K}f_{\phi}$ & 
$F_{K}f_{\phi} $ & 0 &
$-\frac{F_{K}f_{\phi}}{2} $  \\
$\bar{K}^{*0}\phi $ & 0 & 0 & $A_{K^{*}}f_{\phi}$ & $A_{K^{*}}f_{\phi}$ & 
$A_{K^{*}}f_{\phi} $ & 0 &
$-\frac{A_{K^{*}}f_{\phi}}{2}$  \\
\br
\end{tabular}
\end{table}

\begin{table}
\caption{Factorization matrix elements $\tilde{Q}_{i}(h_{1},h_{2})$ 
for $B^{-} \rightarrow h_{1}h_{2}$ decays arising from 
$b \rightarrow s q \bar{q}$.}
\label{table7}
\lineup
\begin{tabular}{@{}llllllll}
\br
Decay  & $\tilde{Q}_{1}$ & $\tilde{Q}_{2}$ & $\tilde{Q}_{3}$ &
$\tilde{Q}_{4}$ & $\tilde{Q}_{5}$ & $\tilde{Q}_{6}$ &
$\tilde{Q}_{9}$ \\
\mr
$\pi^{0}K^{-}$ & $-\frac{F_{\pi}f_{K}}{\sqrt{2}}$ &  
$-\frac{F_{K}f_{\pi}}{\sqrt{2}}$ & 0 & 
$-\frac{F_{\pi}f_{K}}{\sqrt{2}}$ & 0 
&  $-R_{\chi}^{K}\frac{F_{\pi}f_{K}}{\sqrt{2}}$ &
$-\frac{3F_{K}f_{\pi}}{2\sqrt{2}} $ \\
$\pi^{0}K^{*-}$ & $\frac{F_{\pi}f_{K^{*}}}{\sqrt{2}}$ 
& $ \frac{A_{K^{*}}f_{\pi}}{\sqrt{2}}$ & 0 & 
$\frac{F_{\pi}f_{K^{*}}}{\sqrt{2}}$ & 0 & 0 &
$ \frac{3A_{K^{*}}f_{\pi}}{2\sqrt{2}} $ \\
$\pi^{-}\bar{K}^{0}$ & 0 & 0 & 0 & $-F_{\pi}f_{K}$ & 0 
& $-R_{\chi}^{K}F_{\pi}f_{K}$ & 
0  \\
$\pi^{-}\bar{K}^{*0}$ & 0 & 0 & 0 & $F_{\pi}f_{K^{*}}$ & 0 & 0 & 
0  \\
$\rho^{-}\bar{K}^{0}$ & 0 & 0 & 0 & $A_{\rho}f_{K}$ & 0 
& $-R_{\chi}^{K}A_{\rho}f_{K}$ & 
0  \\
$\rho^{-}\bar{K}^{*0}$ & 0 & 0 & 0 & $A_{\rho}f_{K^{*}}$ & 0 & 0 & 
0  \\
$\rho^{0}K^{-}$ & $\frac{A_{\rho}f_{K}}{\sqrt{2}}$ & 
$ \frac{F_{K}f_{\rho}}{\sqrt{2}}$ & 0 &  $\frac{A_{\rho}f_{K}}{\sqrt{2}}$ &
0 & $-R_{\chi}^{K}\frac{A_{\rho}f_{K}}{\sqrt{2}}$ &
$\frac{3F_{K}f_{\rho}}{2\sqrt{2}} $  \\
$\rho^{0} K^{*-}$ & $\frac{A_{\rho}f_{K^{*}}}{\sqrt{2}}$ & 
$\frac{A_{K^{*}}f_{\rho}}{\sqrt{2}}$ & 0 & 
$\frac{A_{\rho}f_{K^{*}}}{\sqrt{2}}$ & 0 & 0 &
$\frac{3A_{K^{*}}f_{\rho}}{2\sqrt{2}} $  \\
$\omega K^{-}$ & $\frac{A_{\omega}f_{K}}{\sqrt{2}}$ & 
$\frac{F_{K}f_{\omega}}{\sqrt{2}}$ & $\sqrt{2}F_{K}f_{\omega}$ & 
$\frac{A_{\omega}f_{K}}{\sqrt{2}}$ & $\sqrt{2}F_{K}f_{\omega}$ &
$-R_{\chi}^{K}\frac{A_{\omega}f_{K}}{\sqrt{2}}$ &
$\frac{F_{K} f_{\omega}}{2\sqrt{2}} $  \\
$\omega K^{*-}$ & $\frac{A_{\omega}f_{K^{*}}}{\sqrt{2}}$ & 
$\frac{A_{K^{*}}f_{\omega}}{\sqrt{2}}$ & $\sqrt{2}A_{K^{*}}f_{\omega} $ &
$ \frac{A_{\omega}f_{K^{*}}}{\sqrt{2}}$ & $\sqrt{2}A_{K^{*}}f_{\omega}$ &
0 & $\frac{A_{K^{*}}f_{\omega}}{2\sqrt{2}} $  \\
$\phi K^{-}$ & 0 & 0 & $F_{K}f_{\phi}$ & $F_{K}f_{\phi}$ & $F_{K}f_{\phi}$ &
0 & $-\frac{F_{K} f_{\phi}}{2} $  \\
$\phi K^{*-}$ & 0 & 0 & $A_{K^{*}}f_{\phi}$ & $A_{K^{*}}f_{\phi}$ &
$A_{K^{*}}f_{\phi}$ & 0 & $-\frac{A_{K^{*}}f_{\phi}}{2}  $  \\
\br
\end{tabular}
\end{table}

\begin{table}
\caption{Measured branching ratio Br(exp), experimental error $\sigma $
(errors added in quadrature), theoretical branching ratio for best
fit parameters Br(fit) and contribution  to $\chi^{2}$ for various $B$
decay channels.}
\label{table8}
\lineup
\begin{tabular}{@{}llllll}
\br
Decay & Br(exp)$^{a}$ & $ \sigma ^{a} $ & Reference$^{b}$
& Br(fit)$^{a}$ & $\chi^{2}$  \\
\mr
$\pi^{+}\pi^{-}$ & 4.1  & 1.2  & Ba1,Be1,Cl1  & 5.4  & 1.07  \\
$\rho^{\pm}\pi^{\mp}$  & 28.9  & 6.9  & Ba2,Be2,Cl1  & 29.6  & 0.01  \\
$\rho^{0}\pi^{0} $ & 3.6  & 3.9  & Ba2  & 0.1  & 0.15  \\
$\omega \pi^{0}$  & 0.8  & 2.0  & Ba3,Cl2  & 0.1  & 0.15  \\
$\rho^{0}\pi^{-}$  & 10.4  & 3.9  & Be2,Cl1  & 9.0  & 0.13  \\
$\omega \pi^{-}$  & 6.6  & 2.2  & Ba3,Cl2  & 7.15  &  0.06  \\
$\pi^{0}K^{-}$  & 10.8  & 2.4  & Ba1,Be1,Cl1  & 11.1  &  0.02  \\
$\pi^{-}K^{0}$  & 18.2  & 3.9  & Ba1,Be1, Cl1  & 18.1  & 0.00  \\
$\pi^{-}K^{*0}$  & 15.9  & 3.7  & Ba4,Be3,Cl2  & 5.0  &  8.75  \\
$\omega K^{-}$  & 3.2  & 2.2  & Cl2  & 1.4  & 0.68  \\
$\phi K^{-} $  & 7.7  & 1.8  &  Ba5,Be2,Cl1  & 7.8  &  0.00  \\
$\phi K^{*-}  $  & 9.6  & 4.4 &  Ba5, Cl1  & 9.6  & 0.00  \\
$\pi^{-}K^{+}$  & 16.7  & 2.2  & Ba1, Be1,Cl1 &  16.9  & 0.01  \\
$\pi^{-}K^{*+} $  & 22  & 11  &  Cl2  & 5.0  & 2.39  \\
$\pi^{0}K^{0} $  & 8.2  & 3.3  & Ba1,Be1,Cl1  & 6.7  & 0.20  \\
$\pi^{0}K^{*0} $  & 2.1  & 2.1  & Cl2  & 1.4  & 0.12  \\
$\omega K^{0}$  & 10  & 6  & Cl2  & 0.7  &  2.42  \\
$\phi K^{0}$  & 8.1  & 3.2  &  Ba5  & 7.4  & 0.05  \\
$\phi K^{*0}$  & 8.6  & 3.0  & Ba5, Be2, Cl1  & 9.0  & 0.02  \\
\br
\end{tabular}\\
\noindent{$^{a}$ In units of $10^{-6}$.}\\
\noindent{$^{b}$ References to experimental groups are:
BaBar Ba1 Ref. \cite{Ba0127}, Ba2 Ref. \cite{Ba0110}, Ba3 Ref. \cite{Ba0115},
Ba4 Ref. \cite{Ba0112}, Ba5 Ref. \cite{Ba0111}; Belle Be1 Ref. \cite{Be0106},
Be2 Ref. \cite{BeBoz}, Be3 Ref. \cite{Be0114} and CLEO Cl1 Ref. \cite{ClGao},
Cl2 Ref. \cite{Cl9913}.}
\end{table}

\begin{table}
\caption{Best fit values and one-standard deviation errors of fitting
parameters.}
\label{table9}
\lineup
\begin{tabular}{@{}lllll}
\br
$ F_{\pi}$ & $F_{K}$ & $A_{\rho} $ & $A_{\omega } $ & $ A_{K^{*}}$ \\
\mr
$0.243 \pm 0.031 $ & $0.312\pm 0.040$ & $0.404\pm 0.087 $ & $ 0.377
\pm 0.067$
& $ 0.349 \pm 0.052 $  \\
&&&& \\
$R^{\pi}_{\chi} $ & $ R^{K}_{\chi}$ & $A$ & $\bar \eta $ & $\bar \rho $  \\
\mr
$0.970 \pm 0.000 $ & $ 1.200 \pm 0.186 $ & $ 0.803 \pm 0.063$ &
$ 0.375 \pm 0.053 $  &  $0.038 \pm 0.104 $  \\
\br
\end{tabular}
\end{table}

\Figures
\begin{figure}
\begin{center}
\epsfsize=0.9\textwidth
\epsfbox{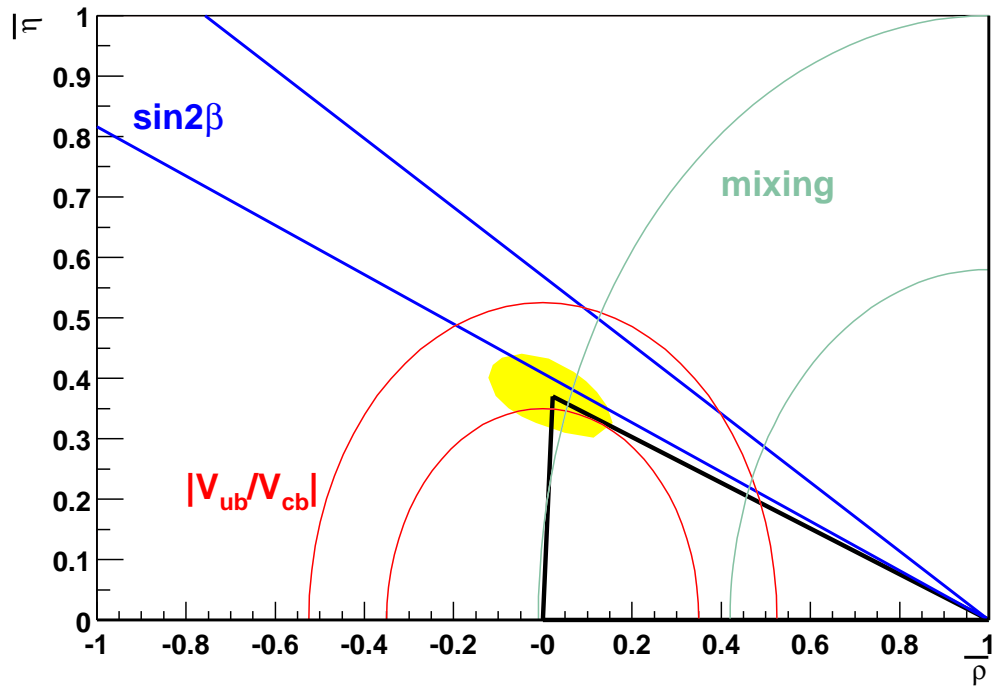}
\end{center}
\caption{The 95\% CL error ellipse for CKM parameters $\bar \rho $ and 
$\bar \eta $ together with the constraints on the Unitarity Triangle from
mixing, $|V_{ub}/V_{cb}|$ and $\sin 2 \beta$.}
\end{figure}

\end{document}